\documentclass[letterpaper, 10 pt, conference]{ieeeconf}  

\IEEEoverridecommandlockouts                              

\usepackage{graphicx} 
\usepackage{hyperref}

\usepackage{cite}
\usepackage{amsmath,amssymb,amsfonts}
\usepackage{algorithmic}
\usepackage{textcomp}
\usepackage{xcolor}
\usepackage{algorithm}

\title{\LARGE \bf
Progressive Curriculum Learning with Scale-Enhanced U-Net for Continuous Airway Segmentation
}

\author{Bingyu Yang, Qingyao Tian, Huai Liao, Xinyan Huang, Jinlin Wu, Jingdi Hu and Hongbin Liu
\thanks{Bingyu Yang and Qingyao Tian are with State Key Laboratory of Multimodal Artificial Intelligence Systems, Institute of Automation, Chinese Academy of Sciences, Beijing 100190, China, and also with the School of Artificial Intelligence, University of Chinese Academy of Sciences, Beijing 100049, China.}%
\thanks{Huai Liao, M.D., Xinyan Huang, M.D. and Jingdi Hu, M.D. are with Department of Pulmonary and Critical Care Medicine, The First Affiliated Hospital of Sun Yat-sen University, Guangzhou, Guangdong Province, P.R. China.}%
\thanks{Jinlin Wu is with Institute of Automation, Chinese Academy of Sciences, Beijing, 100190, China, and also with Centre of AI and Robotics, Hong Kong Institute of Science \& Innovation, Chinese Academy of Sciences.}%
\thanks{Corresponding author: Hongbin Liu is with Institute of Automation, Chinese Academy of Sciences, and with Centre of AI and Robotics, Hong Kong Institute of Science \& Innovation, Chinese Academy of Sciences. He is also affiliated with the School of Biomedical Engineering and Imaging Sciences, King’s College London, UK. (e-mail: liuhongbin@ia.ac.cn).}%
}

\begin{document}

\maketitle
\thispagestyle{empty}
\pagestyle{empty}

\begin{abstract}


Continuous and accurate segmentation of airways in chest CT images is essential for preoperative planning and real-time bronchoscopy navigation.  Despite advances in deep learning for medical image segmentation, maintaining airway continuity remains a challenge, particularly due to intra-class imbalance between large and small branches and blurred CT scan details. To address these challenges, we propose a progressive curriculum learning pipeline and a Scale-Enhanced U-Net (SE-UNet) to enhance segmentation continuity. Specifically, our progressive curriculum learning pipeline consists of three stages: extracting main airways, identifying small airways, and repairing discontinuities. The cropping sampling strategy in each stage reduces feature interference between airways of different scales, effectively addressing the challenge of intra-class imbalance. In the third training stage, we present an Adaptive Topology-Responsive Loss (ATRL) to guide the network to focus on airway continuity. The progressive training pipeline shares the same SE-UNet, integrating multi-scale inputs and Detail Information Enhancers (DIEs) to enhance information flow and effectively capture the intricate details of small airways. Additionally, we propose a robust airway tree parsing method and hierarchical evaluation metrics to provide more clinically relevant and precise analysis. Experiments on both in-house and public datasets demonstrate that our method outperforms existing approaches, significantly improving the accuracy of small airways and the completeness of the airway tree. The code will be released upon publication.

\end{abstract}

\section{INTRODUCTION}

Accurate airway segmentation and reconstruction from CT images are crucial for preoperative planning, real-time navigation, and detecting peripheral pulmonary lesions~\cite{ishiwata2020bronchoscopic}. In clinical practice, only the largest connected component of the segmented airway is used for bronchoscopic localization and navigation to provide a clear and accessible path. This makes continuity in segmentation critical. However, manual segmentation of complex pulmonary airways in CT images is time-consuming and error-prone. Therefore, there is an urgent need for automated technologies to achieve accurate and continuous airway segmentation.


Recent deep learning methods ~\cite{garcia2018automatic,qin2019airwaynet,selvan2020graph, qin2020learning,zheng2021alleviating,nan2023fuzzy,zhang2023towards,wang2024accurate} have made significant strides in improving segmentation accuracy compared to traditional techniques ~\cite{aykac2003segmentation,tschirren2005intrathoracic,kiraly2002three,born2009three}. However, two major challenges remain unaddressed towards achieving continuous airway segmentation: 1) \textit{intra-class imbalance} between major and smaller branches, and 2) \textit{blurred details in CT imaging}. As a result, despite good overlap with ground truth, existing methods often produce fragmented segments, restricting their clinical applicability~\cite{qin2020learning,zheng2021alleviating,nan2023fuzzy,zhang2023towards}.


\textbf{Intra-class Imbalance.} The airway tree has a complex branching structure with significant volume and morphological differences between major and smaller airways, termed intra-class imbalance. Zheng et al.~\cite{zheng2021alleviating} showed that this imbalance causes smaller airway features to be overshadowed by major airways during training, making it difficult for models to focus on smaller structures. Despite their novel supervision approach and distance-based loss function, the issue of airway breakage is not addressed, and missed detections in peripheral small airways persist, leading to airway discontinuities. 
\begin{figure}[!tb]
\centering
\parbox{\columnwidth}{\includegraphics[width=\columnwidth]{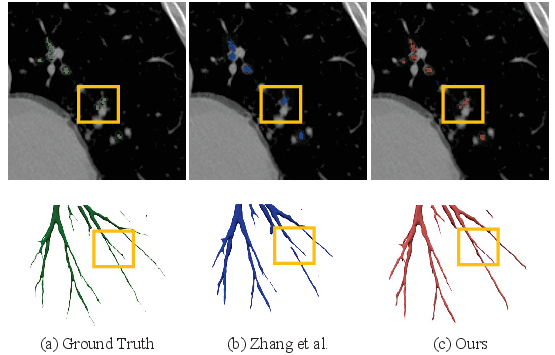}}
\caption{Example of airway discontinuity. (a) Green mask: ground truth. (b) Blue mask: prediction based on Zhang et al.~\cite{zhang2023towards}. (c) Red mask: prediction based on our method. Blurry airway walls lead to discontinuities, which were not effectively addressed by ~\cite{zhang2023towards} (yellow box).}
\label{fig1}
\end{figure}

\textbf{Blurred Details in CT Imaging.} In CT imaging, distal small airways are limited by resolution, noise, and motion blur, often showing low signal-to-noise ratios and blurred details. Fine branches and edge information are easily lost during downsampling, making it hard to restore their integrity during upsampling and causing airway discontinuities. The Connectivity-Aware Surrogate (CAS) and Local-Sensitive Distance (LSD) modules proposed by Zhang et al.~\cite{zhang2023towards} represent the best current approach to addressing airway discontinuities, but they still fail to handle certain blurred airway walls (Fig.~\ref{fig1}).



To address intra-class imbalance and reduce interference in feature learning, we propose a progressive curriculum learning pipeline for airway segmentation, inspired by curriculum learning~\cite{bengio2009curriculum}. We also present a Scale-Enhanced U-Net (SE-UNet) to improve the extraction of airway features across multiple scales, enhancing segmentation accuracy. Additionally, we present hierarchical evaluation metrics to assess performance, emphasizing segmentation continuity at different scales using robust airway tree parsing method.

Motivated by the advancements of curriculum learning in improving segmentation performance by learning from easy to complex tasks~\cite{liu2021style,wang2023grenet,wang2023curriculum}, we propose a progressive curriculum learning pipeline to address intra-class imbalance. This pipeline consists of three stages: extracting main airways, identifying small airways, and repairing discontinuities. To effectively sample training data at each stage, we design three crop sampling strategies (Random Crop, Hard-mining Crop, and Breakage Crop) to mine curriculum samples. A dynamic training scheduler automatically adjusts the input ratio of these samples based on stage-wise performance. To further enhance segmentation continuity, we propose an Adaptive Topology-Responsive Loss (ATRL) to emphasize overlap between predicted and ground-truth centerlines, guiding the network to segment airway with enhanced continuity.


To address discontinuities caused by blurred details in CT imaging, we propose Scale-Enhanced U-Net (SE-UNet). SE-UNet captures airway features at multiple scales, enhancing contextual and edge detail restoration. It uses pyramid pooling on the input image and integrates multi-scale inputs with feature information via residual connections, improving gradient propagation and reducing vanishing gradients. To further enrich critical information, we propose Detail Information Enhancers (DIEs), improving multi-scale feature capture and edge detail recovery.

Existing metrics overlook the evaluation of continuity of segmentation. Moreover, they suffer from bias caused by class imbalance, as larger airways dominate metrics, obscuring deficiencies in small airway segmentation. To accurately evaluate segmentation performance, we present hierarchical evaluation metrics that assess topological completeness at different airway levels. The core is an optimized airway tree parsing algorithm that addresses false branches and handles individual differences by smoothing centerlines, pruning, and merging branches. Extensive experiments on public (499 cases) and in-house datasets (55 cases) demonstrate that our method significantly improves topological completeness compared to state-of-the-art methods, with notable advantages in small airway extraction.

In summary, our contributions are as follows:
\begin{itemize}
\item A progressive curriculum learning pipeline integrated with Adaptive Topology-Responsive Loss (ATRL) is proposed to address intra-class imbalance, enhancing continuous airway segmentation.
\item We propose the Scale-Enhanced U-Net (SE-UNet) with Detail Information Enhancer (DIE) to reduce information loss during downsampling, reducing the impact of blurred details in CT imaging.
\item Hierarchical evaluation metrics are presented to accurately assess segmentation completeness at different scales through a robust airway tree parsing method.
\item Experiments on both public and in-house datasets demonstrate that our method has significant advantages in small airway extraction and airway continuity.
\end{itemize}
\begin{figure*}[!t]
\centering
\parbox{\textwidth}{\includegraphics[width=\textwidth]{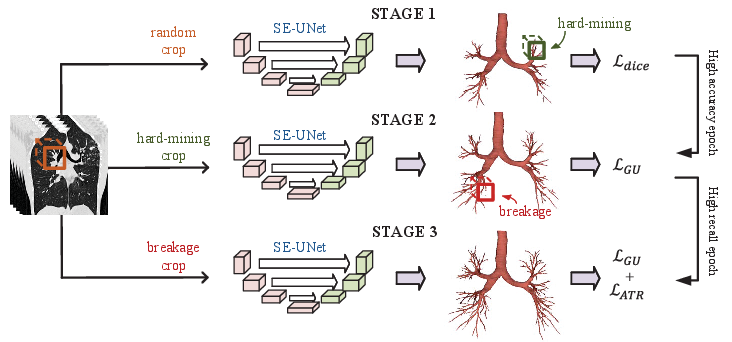}}
\caption{Overview of the proposed progressive curriculum learning pipeline for airway segmentation. In “Stage 1”, the network is trained utilizing the Dice Loss~\cite{milletari2016v}. The epoch with the highest accuracy from the “Stage 1” forms the basis for the hard-mining strategy employed in the “Stage 2”. This stage focuses on learning to detect the challenging airways with GUL~\cite{zheng2021alleviating}. Building upon the predictions with the highest recall from “Stage 2”, the “Stage 3” introduces ATRL to enhance the length of the airway tree. Ultimately, the optimal model is selected for segmentation testing.}
\label{fig2}
\end{figure*}
\begin{figure*}[!tb]
\centering
\parbox{\textwidth}{\includegraphics[width=\textwidth]{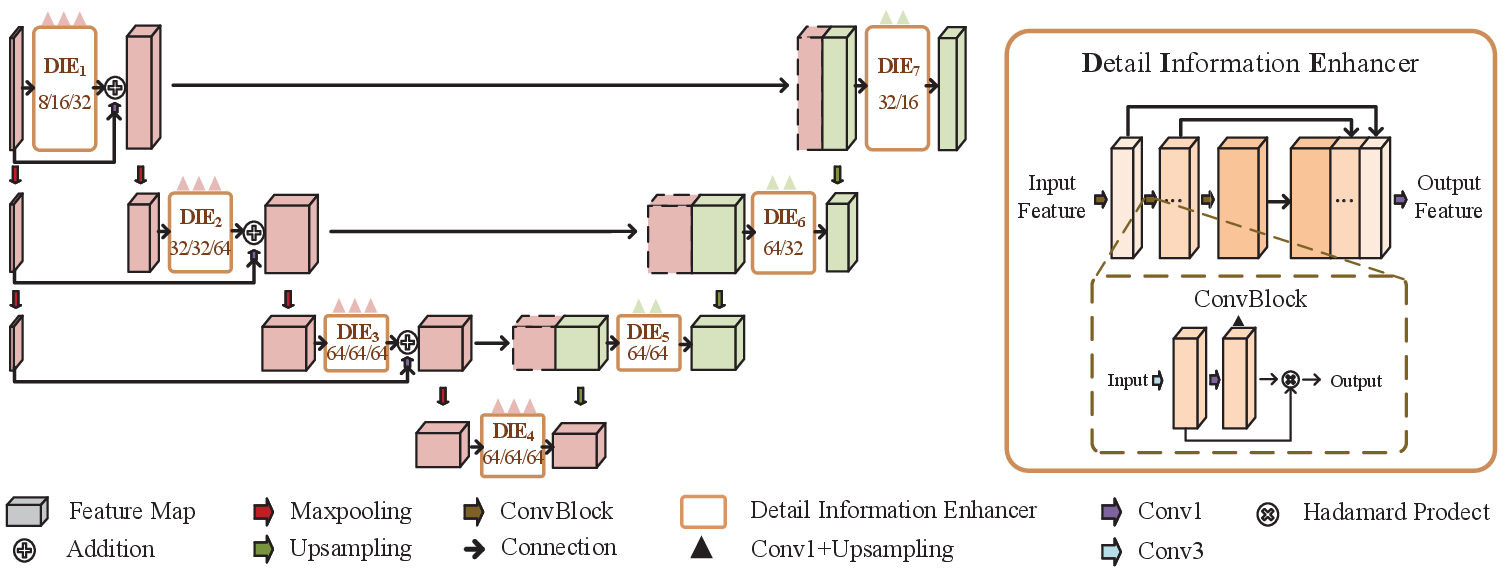}}
\caption{The SE-UNet architecture in proposed method. “$\text{DIE}_1$ $ 8/16/32$” refers to the output channels of the three ConvBlocks in DIE1 are 8, 16 and 32. The detailed structure of DIE is shown in the orange box.}
\label{fig3}
\end{figure*}
\section{RELATED WORKS}

Airway segmentation is crucial for the diagnosis and surgical planning of pulmonary diseases. Traditional methods~\cite{aykac2003segmentation,tschirren2005intrathoracic,kiraly2002three,born2009three,bauer2009segmentation} have been applied. The EXACT 09 challenge~\cite{lo2012extraction} proposed 15 segmentation algorithms. However, the results indicated that these methods are ineffective in extracting small airways.

In recent years, advancements in deep learning have significantly improved the performance of automated airway segmentation~\cite{garcia2018automatic,qin2020learning,selvan2020graph,qin2019airwaynet,zheng2021alleviating,nan2023fuzzy,zhang2023towards}. Juarez et al.~\cite{garcia2018automatic} proposing a 3D U-Net that improved feature representation but struggled with small airway detection. This limitation arose due to CT noise, which caused the features of distal small airways to be blurred.

To enhance feature extraction, Qin et al.~\cite{qin2020learning} added feature recalibration and attention distillation modules, and Selvan et al.~\cite{selvan2020graph} treated the problem as graph refinement. However, the size differences between airway tree branches continue to pose a challenge for feature extraction. Zheng et al.~\cite{zheng2021alleviating} defined this problem as intra-class imbalance, where large airway features dominate the learning process, leading to missed detection of peripheral small airways and resulting in airway discontinuity.

To address intra-class imbalance and airway discontinuity issues, Qin et al.~\cite{qin2019airwaynet} proposed AirwayNet to enhance voxel connectivity. Zheng et al.~\cite{zheng2021alleviating} introduced the General Union Loss (GUL) to alleviate this problem. The distance-based loss function improved the model’s focus and learning of the airway centerline but did not directly solve the airway discontinuity problem. Zhang et al.~\cite{zhang2023towards} introduced Connectivity-Aware Surrogate (CAS) and Local-Sensitive Distance (LSD) modules, which improved connectivity but led to over-segmentation issues.

In this work, we propose a progressive curriculum learning pipeline based on curriculum learning that effectively addresses the intra-class imbalance issue, and a SE-UNet which enhances multi-scale information.

\section{METHOD}


In the following section, we present the progressive curriculum learning pipeline (\S III.A) and SE-UNet (\S III.B), which enhances information flow and segmentation details. Additionally, we propose hierarchical evaluation metrics to provide fair evaluation emphasizing continuity (\S III.C).

\subsection{Progressive curriculum learning pipeline} 

Due to the large volume of 3D lung CT data and GPU memory limitations, airway segmentation networks are trained using sampling patches as inputs~\cite{garcia2018automatic,nan2023fuzzy}. When patches contain both large and small airways, intra-class imbalance causes loss gradients to be dominated by large airways, exacerbating gradient erosion in small airways~\cite{zheng2021alleviating}. To address this, we design a progressive curriculum learning pipeline to train the network and minimize interference during feature learning (Fig.~\ref{fig2}). The pipeline consists of three stages: extracting main airways, identifying small airways, and repairing discontinuities. We develop three crop sampling strategies to effectively mine curriculum samples, and the training scheduler dynamically adjusts the input ratio of these samples based on the performance of each stage. The specific design and function of each stage are as follows:

\textbf{Stage 1:} The network is trained on airway patches using Random Crop Sampling to learn basic tubular features and achieves rapid convergence under the supervision of Dice loss~\cite{milletari2016v}:
$$
\mathcal{L}_{1}=\mathcal{L}_{dice}=1-\frac{2\sum_{i=1}^{N}p_{i}g_{i}}{\sum_{i=1}^{N}(p_{i}+g_{i})}, \eqno{(1)}
$$
where $N$ is the total number of voxels, $p_{i}$ is the i-th voxel' prediction and $g_{i}$ is the ground truth. 

\textbf{Stage 2:} Hard-mining Crop Sampling is introduced to select unextracted airway skeleton points from Stage 1 predictions to generate patches. The training scheduler dynamically adjusts the weights based on the quantities of random and hard-mining sampling patches. This stage is supervised by General Union Loss (GUL)~\cite{zheng2021alleviating} to address potential intra-class imbalance and enhance the extraction of peripheral small airways:
$$
\mathcal{L}_{2}=\mathcal{L}_{GU}=1-\frac{\sum_{i=1}^{N}w_{l}p_{i}^{\gamma}g_{i}}{\sum_{i=1}^{N}w_{l}(\alpha p_{i}+\beta g_{i})}, \eqno{(2)}
$$
where Local-imbalance-based Weight $w_{l}$~\cite{zheng2021refined} adjusts airway voxel weights by branch size. According to~\cite{zheng2021alleviating}, we choose $\gamma=0.7$, $\alpha=0.2$, $\beta=0.8$.

\begin{figure}[!tb]
\centering
\parbox{\columnwidth}{\includegraphics[width=\columnwidth]{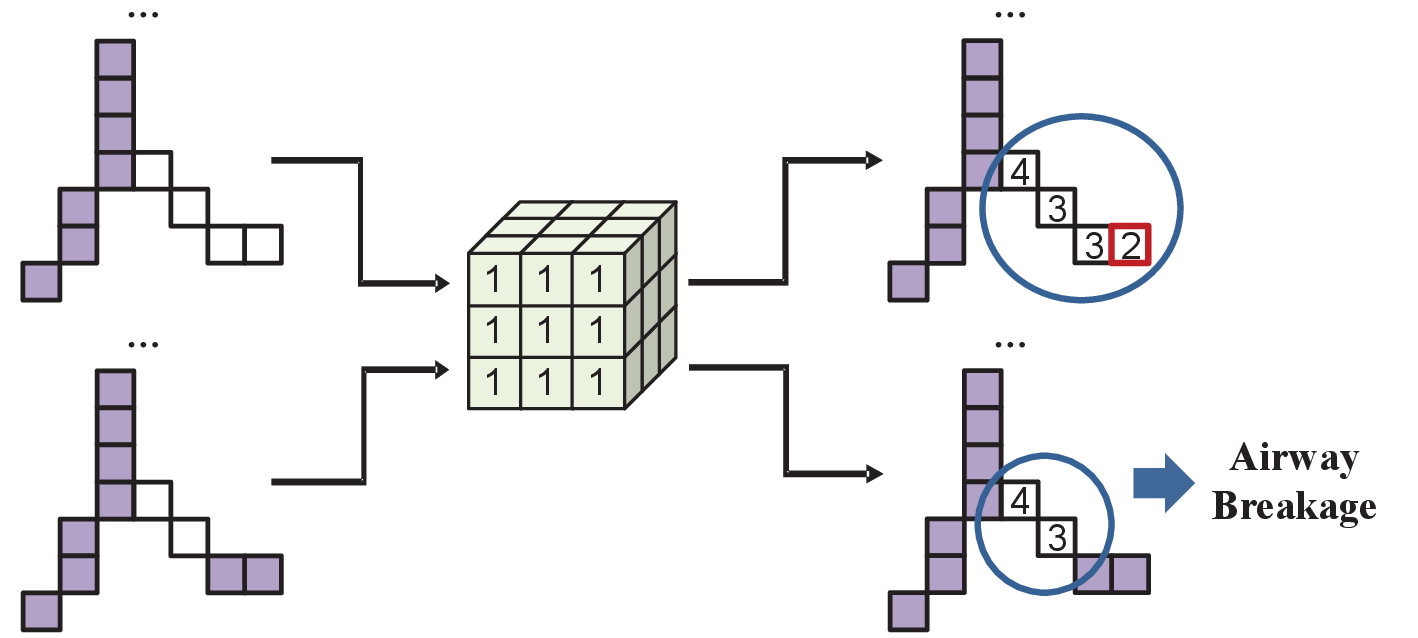}}
\caption{Explanation of identifying airway breakages. Here shows an airway skeleton segment, with purple points representing successful predictions and white points indicating failures. A 3×3×3 all-ones convolution kernel is applied to the white points to count their neighbors. If no point on a challenging airway segment has a single neighbor, it is identified as a breakage.}
\label{fig4}
\end{figure}

\textbf{Stage 3:} The most challenging aspect is optimizing airway discontinuities heavily influenced by CT noise. We propose Breakage Crop Sampling to isolate discontinuity-related segments using a 3×3×3 all-ones convolution kernel on challenging airway skeleton points from Stage 2 (as shown in Fig.~\ref{fig4}). Similarly, the training scheduler dynamically adjusts the ratio of random, hard-mining, and breakage sampling patches. In this stage, we propose a Adaptive Topology-Responsive Loss (ATRL), as defined in (3). This loss function penalizes voxels that are difficult to detect within the airway skeleton, placing greater emphasis on the continuity of the airway.
$$
\mathcal{L}_{ATR}=1-\frac{\sum_{i=1}^{C}w_{i}p_{i}g_{i}}{\sum_{i=1}^{C}w_{i}(p_{i}+g_{i})}, \eqno{(3)}
$$
where $C$ is the number of voxels on the airway centerline. To focus on hard-to-segment small airways, $w_{i}$ is is written as follow:
$$
w_{i}=w_{l}+w_{c}, \eqno{(4)}
$$
where Centerline-distance-based Weight $w_{c}$ provides increased attention to points near the difficult-to-segment airway centerline to maintain branch continuity, with additional weight assigned to voxels that contribute to breakage:
$$
w_{c}=\left(1-\frac{d_{i}}{d_{max}}\right)^{2}+\eta \cdot min(1-d_{i},K),\eqno{(5)}
$$
where $d_{i}$ is the shortest distance from the i-th voxel to the centerline, $d_{max}$ is its maximum value. $\eta$ indicates whether the i-th voxel induces breakage ($\eta$=1 if it does, otherwise $\eta$=0). $K$ is a hyper-parameter set to 2 to prevent $w_{c}$ from becoming too large.

In the third stage of training, the hybrid loss function is:
$$
\mathcal{L}_{3}=\mathcal{L}_{GU}+\mathcal{L}_{ATR}.\eqno{(6)}
$$

\subsection{Scale-Enhanced U-Net}

The pulmonary airway structure is complex, consisting of multi-scale airway branches that pose significant challenges in feature extraction due to imaging noise and intra-class imbalance. To address these issues, we propose a Scale-Enhanced U-Net (SE-UNet). This architecture achieves hierarchical residual nesting through the use of multi-scale inputs and Detail Information Enhancers (DIEs). The overall network structure is depicted in Fig.~\ref{fig3}. Specifically, the 3D patches input to the network are pyramid-pooled into three resolutions, with residuals computed using a 1×1×1 convolution layer and the output features from the residual multi-scale encoder at each stage. The SE-UNet effectively captures and integrates multi-scale airway information during the encoding phase, enhancing its ability to model complex airway structures.

The SE-UNet comprises a total of 7 Detail Information Enhancers (DIEs), divided into an encoding group (DIE 1-4) and a decoding group (DIE 5-7). In encoder, the features extracted by the i-th layer are obtained by:
$$
F_{en_{i}}=\mathrm{Conv1}(\mathrm{Maxpool}_{i}({F_{in}}))\oplus F_{DIE_{i}}, \eqno{(7)}
$$
where $\mathrm{Conv1}$ is 1×1×1 convolution, $\mathrm{Maxpool}_{i}$ represents the maxpooling at the i-th depth level and $\oplus$ denotes addition. $F_{in}$ is the input feature and $F_{DIE_{i}}$ is the output of the DIE at the i-th encoder layer.

Similarly, the features recovered by the i-th decoding layer are obtained by:
$$
F_{de_{i}}=F_{DIE_{i}}. \eqno{(8)}
$$
Group supervision has been shown to effectively alleviate inter-class imbalance~\cite{zheng2021alleviating}, so it is introduced for supervising the encoding and decoding groups in SE-UNet.

\textbf{Detail Information Enhancer:} The DIE is an Inception-like architecture~\cite{szegedy2015going} that integrates multi-scale features extracted by convolutional layers of varying depths and enhances details, as shown in the orange box of Fig.~\ref{fig3}. Each DIE consists of $n$ ConvBlocks, “$\text{DIE}_i$ $\text{c}_{1}/\ldots/\text{c}_{n}$” refers to the output channels of $n$ ConvBlocks in $\text{DIE}_i$. Each ConvBlock includes convolution layers, an instance normalization layer and a ReLU layer. It outputs features to the next block $f_{i,j}$ while generating an additional path through the upsampling layer, forming a group feature pyramid that is used for group supervision~\cite{zheng2021alleviating}. The output of the i-th DIE to the next module $F_{DIE_{i}}$ is calculated as:
$$
F_{DIE_{i}}=\mathrm{Conv1}(\mathrm{Cat}(f_{i,1},f_{i,2},\ldots,f_{i,n})), \eqno{(9)}
$$
where $\mathrm{Cat}$ denotes concatenation, $f_{i,j}$ represents the feature of the j-th ConvBlock in the i-th DIE. $n$ is the total number of blocks in this DIE (3 in the encoder and 2 in the decoder).
$$
f_{0,j}=\mathrm{RELU}(\mathrm{IN}(\mathrm{Conv3}(f_{i,j-1}))), \eqno{(10)}
$$
$$
f_{i,j}=\mathrm{Sigmoid}(\mathrm{Conv1}(f_{0,j}))\odot f_{0,j}, \eqno{(11)}
$$
where $f_{0,j}$ is a intermediate result, $\mathrm{IN}$ denotes instance normalization, $\mathrm{Conv3}$ is 3×3×3 convolution, $f_{i,j-1}$ represents the output of the previous block. $\odot$ is Hadamard product.

\subsection{Hierarchical Evaluation}
\begin{figure}[!tb]
\centering
\parbox{\columnwidth}{\includegraphics[width=\columnwidth]{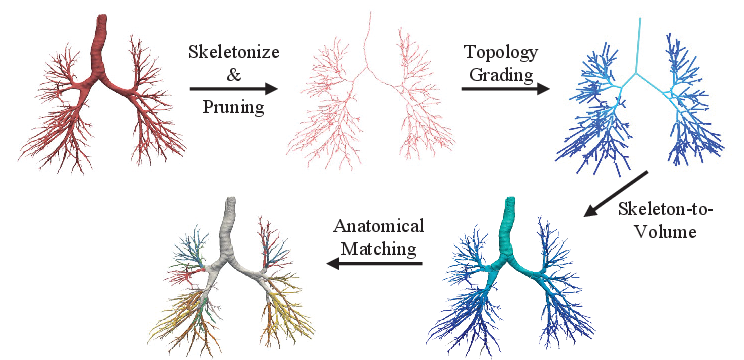}}
\caption{A qualitative example illustrating the topological analysis process of airway tree.}
\label{fig5}
\end{figure}

Current airway segmentation research uses Tree length Detected rate(TD) and Branch Detected rate(BD) to assess airway completeness, Dice Similarity Coefficient(DSC) and Precision(Pre) to measure accuracy~\cite{zhang2023multi}. However, these metrics are limited by false branches and intra-class imbalance, masking deficiencies in small airway segmentation. Thus, we propose hierarchical evaluation metrics based on Robust Airway Tree Parsing (Fig.~\ref{fig5}) to evaluate topological completeness at different airway levels.

\textbf{Robust Airway Tree Parsing:} Existing methods include skeleton-based analysis~\cite{palagyi2003quantitative} and deep learning-based classification~\cite{tan2021sgnet}. However, the former often generates inaccurate skeletons, while the latter misses high-order branches. We propose a robust airway tree parsing pipeline that includes Skeletonization and Parsing, Pruning, Topology Grading, and Anatomical Matching. (a) Skeletonization and Parsing: Traverse the extracted skeleton to identify branch points and explore along the skeleton line from each branch point until the airway end or the next branch point is reached. Record parent-child relationships to complete skeleton parsing. (b) Pruning: Smooth the main airway segments, remove peripheral short segments, and merge single-segment parent branches to accurately divide airway branches. (c) Topology Grading: Recursively number branch points along the skeleton. (d) Anatomical Matching: Align the original numbering with anatomical grades based on hierarchical relationships and airway angles, considering common branch omissions and individual differences to achieve stable airway tree parsing.


\textbf{Evaluation Metrics:} 
In addition to the commonly used metrics (TD, BD, DSC, and Pre), we propose hierarchical evaluation metrics to mitigate biases caused by the intra-class imbalance. Following the anatomical labeling in~\cite{tschirren2005matching}, we define two levels: Large airways = trachea + main bronchi + lobar bronchus (gray airways after anatomy matching in Fig.~\ref{fig5}) and Small airways = segmental airways (colored airways). The hierarchical evaluation metrics for airway completeness are formulated as follows:
$$
TD_{L(S)}=\frac{\hat{T}_{L(S)}}{T_{L(S)}}, \eqno{(12)}
$$
$$
BD_{L(S)}=\frac{\hat{B}_{L(S)}}{B_{L(S)}}, \eqno{(13)}
$$
where $\hat{T}_{L(S)}$ refers to the total detected branch length for Large (Small) airways, while $T_{L(S)}$ represents the length of the these branches in the ground truth. $\hat{B}_{L(S)}$ denotes the number of correctly detected Large (Small) branches, and $B_{L(S)}$ is the ground truth.

\section{EXPERIMENT}

\subsection{Dataset}

We evaluate our method on two datasets: in-house dataset used for training, assessment, and ablation studies; the ATM’22 challenge dataset~\cite{zhang2023multi} for fair comparative testing\footnote{\url{https://atm22.grand-challenge.org/}}.

\textbf{ATM’22 challenge dataset:} This dataset provides 499 CT scans: 299 for training, 50 for validation, and 150 for testing~\cite{qin2019airwaynet,yu2022break,zhang2022cfda,zheng2021alleviating}, sourced from EXACT’09~\cite{lo2012extraction}, LIDC-IDRI~\cite{armato2011lung}, and Shanghai Chest Hospital. Three radiologists with over five years of experience meticulously annotated the CT images. We trained our model on the train dataset, with test results evaluated by the challenge organizers.

\textbf{In-house dataset:} We collected and annotated 55 CT scans from The First Affiliated Hospital of Sun Yat-sen University, including samples from healthy subjects as well as patients with mild COPD and ILD. Each scan contains 256 to 612 slices, with spatial resolution ranging from 0.598 to 0.879 mm and slice thickness from 0.625 to 1.000 mm. The scans were split into training (35), validation (10), and test sets (10). Initial segmentation was done using models~\cite{zheng2021alleviating}, followed by detailed annotations by three pulmonologists, finalized through majority voting.
\begin{figure*}[!tb]
\centering
\parbox{\textwidth}{\includegraphics[width=\textwidth]{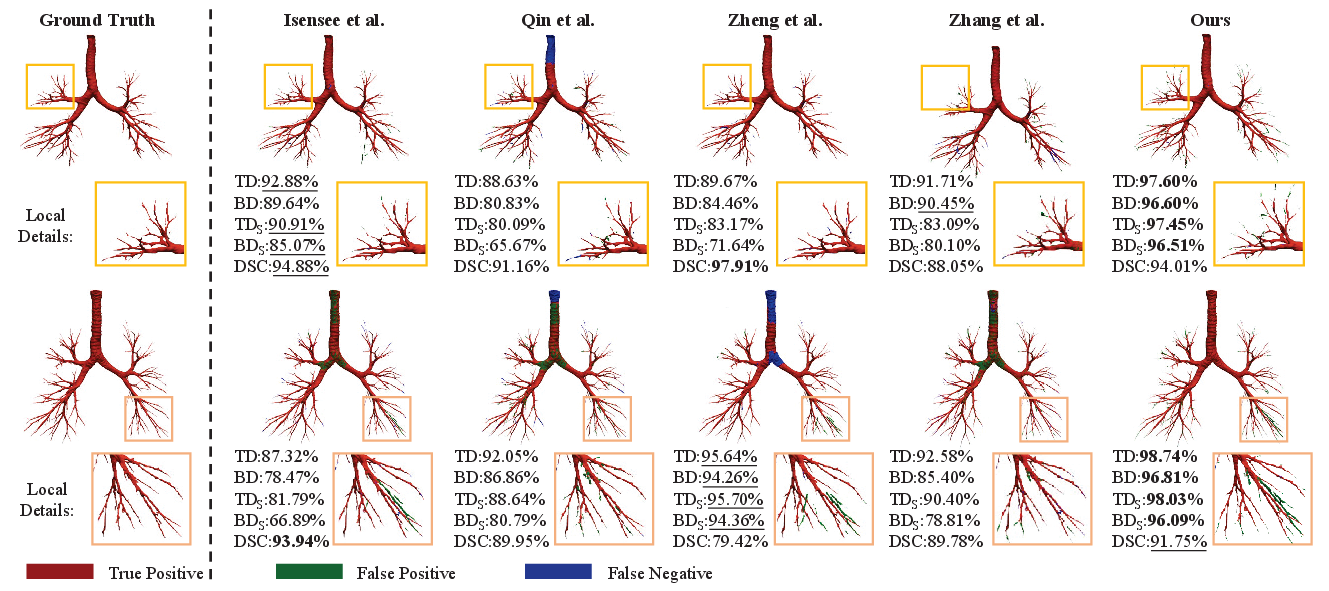}}
\caption{Visualization of the segmentation results of the our approach and other SOTA methods. Red, green, and blue represent correct airways(true positive), airway leakages(false positive) and missing airways(false negative), respectively. Best viewed in color and magnified.}
\label{fig6}
\end{figure*}
\subsection{Implementation Details}

Preprocessing involves truncating CT voxel intensities to [-1000,500] and [-1024,1024], followed by normalization to [0,1]. The network is trained using 128×128×128 CT patches as input. The optimizer used is AdamW with a batch size of 2 and a learning rate of 0.0001. The training epochs for the three stages are set to 100, 50, and 50, respectively. For post-processing, Dual Threshold Iteration (DTI)~\cite{liu2022full} is employed to convert the probability map to a binary mask, followed by morphological hole filling and extraction of the largest connected component. Thresholds in DTI are set at $T_{h}=0.5$ and $T_{l}=0.35$. The framework is implemented using PyTorch 2.0.0 and both training and inference are conducted on a single NVIDIA RTX 4090 GPU. In inference, our approach can achieve an average segmentation time of 73 ms per 128×128×128 CT patch.

\section{RESULTS}

\subsection{Comparison Results}
\begin{table}[!tb]
    \centering
    \caption{Comparison on ATM22 test dataset. The \textbf{best} and \underline{second-best} are highlighted.}\label{tab1}
    \renewcommand\arraystretch{1.1} 
    \setlength{\tabcolsep}{1.8mm}{
        \begin{tabular*}{\hsize}{@{}@{\extracolsep{\fill}}llllllll@{}}
        \hline
            Team & TD(\%) & BD(\%) & DSC(\%) & Pre(\%) & wMS  \\ \hline            
            neu204~\cite{wu2023two} & 90.974  & 86.670  & 94.056  & 93.027 &  90.710 \\ 
            ~ & \textcolor{gray}{±10.409} & \textcolor{gray}{±13.087} & \textcolor{gray}{±8.021} & \textcolor{gray}{±8.410} \\ 
            YangLab~\cite{nan2023fuzzy} & 94.512  & 91.920  & 94.800  & 94.707 &  93.831   \\ 
            ~ & \textcolor{gray}{±8.598} & \textcolor{gray}{±9.435} & \textcolor{gray}{±7.925} & \textcolor{gray}{±8.302} \\ 
            deeptree\_damo~\cite{wang2024accurate} & 97.853  & 97.129  & 92.819  & 87.928 &  94.644  \\ 
            ~ & \textcolor{gray}{±2.275} & \textcolor{gray}{±3.411} & \textcolor{gray}{±2.191} & \textcolor{gray}{±4.181}  \\ 
            timi & 95.919  & 94.729  & 93.910  & 93.553 &  \underline{94.687}    \\ 
            ~ & \textcolor{gray}{±5.234} & \textcolor{gray}{±6.385} & \textcolor{gray}{±3.682} & \textcolor{gray}{±3.420} \\ 
            Ours & 96.425 & 95.479 & 93.827 & 91.781 &  \textbf{94.693}  \\ 
            ~ & \textcolor{gray}{±3.156} & \textcolor{gray}{±4.313} & \textcolor{gray}{±1.897} & \textcolor{gray}{±4.110} \\
            \hline
        \end{tabular*}}
\end{table}
The ATM'22 challenge organizers conducted a quantitative analysis of our method's performance on a hidden test set. Table~\ref{tab1} compares our test results with those of the top four teams in the competition, ranked by weighted mean score(wMS). To assess the sensitivity of different algorithms to airway continuity, we emphasize the geometric shape of airway predictions through the weighted average score. The weighted mean score(wMS) is calculated as 30\% TD, 30\% BD, 20\% DSC and 20\% Precision. Our method achieved the highest weighted average score of 94.693. Notably, our method outperformed timi in TD (96.425\%) and BD (95.479\%) scores. Our precision score (91.781\%) was slightly lower than other metrics. After a careful analysis of the training and validation sets provided by the organizers by medical experts, it was found that some peripheral branches in the dataset were not fully annotated. We speculate that this might be because our method successfully detected these unannotated peripheral branches on the hidden test set, thereby affecting the precision metric. As illustrated in Fig.~\ref{fig6}, our algorithm exhibits a reliable capacity for detecting more peripheral airways. Compared to other teams, our TD and BD scores were 1.9\% and 3.6\% higher than the YangLab team~\cite{nan2023fuzzy}, and 5.5\% and 8.8\% higher than the neu204 team~\cite{wu2023two}. The deeptree\_damo team~\cite{wang2024accurate}, in their pursuit of higher airway completeness, sacrificed precision, resulting in the lowest DSC (92.819\%) and Precision scores (87.928\%). Overall, our method strikes an excellent balance between airway accuracy and completeness, achieving the highest mean score.
\begin{table}[!tb]
    \centering
    \caption{Comparison on in-house dataset.$^\ast$ denotes results from an open-source module (with trained weights). $^\dagger$ indicates the results from modules trained using open-source codes.}\label{tab2}
    \renewcommand\arraystretch{1.1} 
    \setlength{\tabcolsep}{3mm}{
        \begin{tabular*}{\hsize}{@{}@{\extracolsep{\fill}}llllllll@{}}
        \hline
            Method & TD(\%) & BD(\%) & DSC(\%) & Pre(\%)  \\ \hline
            
            
            
            Isensee et al.~\cite{isensee2021nnu}$^\dagger$ & 82.453  & 72.019  &  \textbf{93.132} & \textbf{92.931}  \\ 
            ~ & \textcolor{gray}{±11.817} & \textcolor{gray}{±14.990} & \textbf{\textcolor{gray}{±4.065}} & \textbf{\textcolor{gray}{±7.323}} \\ 
            
            Qin et al.~\cite{qin2021learning}$^\ast$ & 86.300  & 81.783  & 88.755  & 89.870   \\ 
            ~ & \textcolor{gray}{±11.571} &\textcolor{gray}{±15.655} & \textcolor{gray}{±3.373} & \textcolor{gray}{±4.340} \\ 
            
            Zheng et al.~\cite{zheng2021alleviating}$^\ast$ & 89.296  & 84.739  & 91.052  & \underline{90.814} \\ 
            ~ & \textcolor{gray}{±2.872} & \textcolor{gray}{±4.389} & \textcolor{gray}{±1.098} & \underline{\textcolor{gray}{±6.769}}   \\ 
            
            Zhang et al.~\cite{zhang2023towards}$^\ast$ & \underline{90.434}  & \underline{85.418}  & 88.803  & 81.407 \\ 
            ~ & \underline{\textcolor{gray}{±3.060}} & \underline{\textcolor{gray}{±6.807}} & \textcolor{gray}{±0.760} & \textcolor{gray}{±1.154}   \\ 
                 
            Ours & \textbf{92.784}  & \textbf{90.040}  & \underline{91.488}  & 87.190 \\ 
            ~ & \textbf{\textcolor{gray}{±3.665}} & \textbf{\textcolor{gray}{±5.7327}} & \underline{\textcolor{gray}{±1.230}} & \textcolor{gray}{±2.322} \\ \hline

             & $\text{TD}_L$(\%) & $\text{BD}_L$(\%) &  $\text{TD}_S$(\%) &  $\text{BD}_S$(\%)  \\ \hline
            Isensee et al.~\cite{isensee2021nnu}$^\dagger$ & 87.324 & 82.775 & 70.425 & 59.375\\ 
            Qin et al.~\cite{qin2021learning}$^\ast$ & 90.820  & 88.536  & 70.721  & 61.924   \\ 
            Zheng et al.~\cite{zheng2021alleviating}$^\ast$ & 91.862  & 91.457  & 75.545  & 65.407 \\ 
            Zhang et al.~\cite{zhang2023towards}$^\ast$ & \underline{93.350}  & \underline{91.493}  & \underline{80.903}  & \underline{71.710} \\ 
            Ours & \textbf{94.857}  & \textbf{92.452}  & \textbf{90.534}  & \textbf{74.532}  \\ \hline
            
        \end{tabular*}}
\end{table}

Based on the proposed robust airway tree parsing, the evaluation results on our in-house dataset are shown in Table~\ref{tab2}. Our method is compared with other state-of-the-art methods from the past five years, achieving the highest TD (92.784\%), BD (90.040\%) and the second-highest DSC at 91.488\%. The nnUNet by Isensee et al.~\cite{isensee2021nnu} achieved the highest DSC (93.132\%) and Precision (92.931\%), but it does not specifically address airway segmentation challenges, resulting in poorer airway continuity. Qin et al.~\cite{qin2021learning} combined a 3D UNet with an attention distillation module to improve the detection of peripheral small airways, achieving TD (86.300\%) and BD (81.783\%). However, their method did not effectively address the issue of class imbalance. Zheng et al.~\cite{zheng2021alleviating} introduced the GUL and a new supervision to tackle class imbalance. However, GUL did not directly penalize airway discontinuities, resulting in TD (89.296\%) and BD (84.739\%). Zhang et al.~\cite{zhang2023towards} proposed a Connectivity-Aware Surrogate (CAS) module to improve airway continuity but introduced potential over-segmentation issues. Although their method achieved the second-highest TD (90.434\%) and BD (85.418\%), it had the lowest Pre at 81.407\%. To compare the completeness across different airway branch sizes, the lower part of Table~\ref{tab2} shows the tree length and branch detection results for large airways (trachea, main bronchi, lobar bronchus) and small airways (segmental airways), denoted as $\text{TD}_L$, $\text{BD}_L$ and $\text{TD}_S$, $\text{BD}_S$. Fig.~\ref{fig6} visualizes the segmentation results of our method and other SOTA methods, further demonstrating our significant advantages in small airway segmentation and airway tree continuity.

\subsection{Ablation Studies}

In our in-house dataset, we conducted a detailed analysis of our method's components: (a) the effectiveness of the Detail Information Enhancer (DIE); (b) the impact of the Adaptive Topology-Responsive Loss (ATRL); and (c) the importance of the second and third stages in the progressive curriculum learning pipeline. Table~\ref{tab3} presents the quantitative results of this ablation study.
\begin{table}[!tb]
    \centering
    \caption{Ablation study on in-house dataset. ‘DIE’:Detail Information Enhancer; ‘DIE\_{en}’or ‘DIE\_{de}’:DIE in the encoder or decoder; ‘ATRL’:Adaptive Topology-Responsive Loss.}\label{tab3}
    \renewcommand\arraystretch{1.1} 
    \setlength{\tabcolsep}{3mm}{
        \begin{tabular*}{\hsize}{@{}@{\extracolsep{\fill}}llllllllll@{}}
        \hline
            Method & TD(\%) & BD(\%) & DSC(\%) & Pre(\%)   \\ \hline
            
            Ours w/o DIE & 88.358  & 82.400  & 91.868  & 87.676    \\ 
            ~ & \textcolor{gray}{±5.787} & \textcolor{gray}{±9.767} & \textcolor{gray}{±0.530} & \textcolor{gray}{±1.105}   \\  
            
            Ours w/o DIE\_{en} & 89.066  & 83.396  & 93.038  & \underline{89.563}   \\ 
            ~ & \textcolor{gray}{±5.363} & \textcolor{gray}{±8.110} & \textcolor{gray}{±1.025} & \underline{\textcolor{gray}{±2.013}} \\    
            
            Ours w/o DIE\_{de} & 89.529  & 83.850  & 92.918  & 89.121   \\ 
            ~ & \textcolor{gray}{±5.940} & \textcolor{gray}{±8.718} & \textcolor{gray}{±0.472} & \textcolor{gray}{±1.091} \\ \hline
            
            Ours w/o ATRL & \underline{90.118}  & \underline{84.758}  & \textbf{93.244}  & 89.502     \\ 
            ~ & \underline{\textcolor{gray}{±4.674}} & \underline{\textcolor{gray}{±7.846}} & \textbf{\textcolor{gray}{±0.534}} & \textcolor{gray}{±1.207}  \\  \hline
            
            Ours\_Stage1 & 71.922  & 60.286  & \underline{93.118}  & \textbf{91.593}  \\ 
            ~ & \textcolor{gray}{±7.993} & \textcolor{gray}{±8.419} &\underline{ \textcolor{gray}{±0.867}} & \textbf{\textcolor{gray}{±0.941}}  \\ 

            Ours\_Stage1+2 & 88.068  & 81.550  & 92.908  & 89.309   \\ 
            ~ & \textcolor{gray}{±5.734} & \textcolor{gray}{±8.514} & \textcolor{gray}{±0.943} & \textcolor{gray}{±1.752}  \\ \hline
            Ours & \textbf{92.784}  & \textbf{90.040}  & 91.488  & 87.190    \\ 
            ~ & \textbf{\textcolor{gray}{±3.665}} & \textbf{\textcolor{gray}{±5.733}} & \textcolor{gray}{±1.230} & \textcolor{gray}{±2.322}   \\ \hline
        \end{tabular*}
    }
\end{table}

\begin{figure}[!tb]
\centering
\parbox{\columnwidth}{\includegraphics[width=\columnwidth]{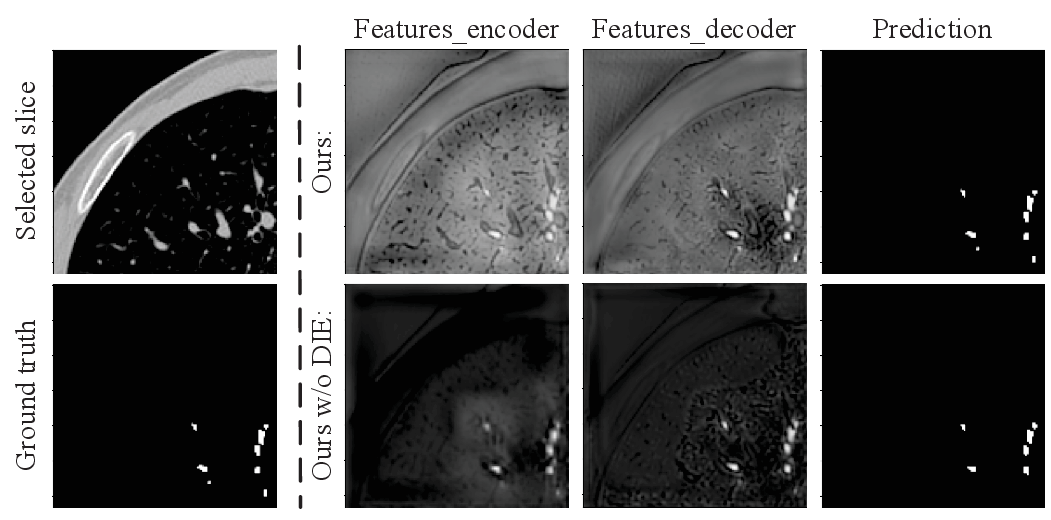}}
\caption{The impact of the DIE. The feature maps from the encoder and decoder are shown on the right.}
\label{fig7}
\end{figure}

Firstly, we evaluated the impact of removing the Detail Information Enhancers (DIEs). Omitting DIEs (Ours w/o DIE) slightly affected the DSC and Pre but significantly reduced TD and BD by 4.426\% and 7.640\%, respectively. This indicates that DIEs play a crucial role in maintaining airway continuity, especially when segmenting the blurred airways in CT images (as shown in Fig.~\ref{fig7}). Specifically, the absence of multi-scale inputs and DIEs in the encoding stage (Ours w/o DIE\_{en}) notably impaired gradient propagation for small airways, thereby affecting the integrity of airway prediction. In contrast, DIEs in the decoder successfully restored the blurred features, significantly improving tree length.

Secondly, the effectiveness of ATRL was explored. The model incorporating ATRL showed improvements in TD (2.666\%) and BD (5.282\%), confirming that ATRL effectively encourages the network to focus on the airway breakages, thereby enhancing the segmentation accuracy.

Lastly, we compared the network trained through all three stages with those trained through only the first stage or the first two stages, with the results shown in the last three rows of Table~\ref{tab3}. The second and third stages improved TD and BD by 16.146\% and 4.716\%, and by 21.264\% and 8.490\%, respectively. It shows that the second stage enhances peripheral airway detection, while the third stage further improves airway continuity by addressing discontinuities.

\section{CONCLUSIONS}
In this study, we propose a progressive curriculum learning pipeline and a Scale-Enhanced U-Net (SE-UNet), achieving accurate and continuous airway segmentation. The progressive curriculum learning pipeline reduces feature interference between different scales and alleviates intra-class imbalance. Additionally, the Adaptive Topology-Responsive Loss (ATRL) proposed in the third stage further optimizes airway continuity. All three training stages share the same Scale-Enhanced U-Net (SE-UNet), which captures the blurred details of small airways through multi-scale inputs and Detail Information Enhancers (DIEs). Experiments demonstrate that our method outperforms existing approaches in small airway detection and airway tree completeness, providing more reliable segmentation results for preoperative planning and real-time bronchoscopy navigation.

\bibliographystyle{IEEEtran}
\bibliography{IEEEabrv,ref}

\end{document}